\documentclass[aps,pre,twocolumn,showpacs]{revtex4}

\usepackage{graphicx}
\usepackage{amssymb}
\usepackage{amsmath}

\newcommand{\pd}{p_{\downarrow\downarrow}}
\newcommand{\pu}{p_{\downarrow\uparrow}}
\newcommand{\pperp}{p_{\perp}}
\newcommand{\df}{d_f}
\newcommand{\Pe}{{\rm Pe}}
\newcommand{\ave}[1]{\left\langle#1\right\rangle}

\begin{document}

\title{No self-similar aggregates with sedimentation}

\author{M. Peltom\"aki}
\email{ppv@fyslab.hut.fi}
\affiliation{Laboratory of Physics, Helsinki University of Technology,
P. O. Box 1100, FIN-02150 HUT, Finland}
\author{E. K. O. Hell\'en}
\email{ehe@fyslab.hut.fi}
\affiliation{Laboratory of Physics, Helsinki University of Technology,
P. O. Box 1100, FIN-02150 HUT, Finland}
\author{M. J. Alava}
\email{mja@fyslab.hut.fi}
\affiliation{Laboratory of Physics, Helsinki University of Technology,
P. O. Box 1100, FIN-02150 HUT, Finland}
\affiliation{SMC-INFM, Dipartimento di Fisica,
Universit\`a ``La Sapienza'', P.le A. Moro 2
00185 Roma, Italy}
\date{\today}

\begin{abstract}

Two-dimensional cluster-cluster aggregation is studied when clusters
move both diffusively and sediment with a size dependent
velocity. Sedimentation breaks the rotational symmetry and the ensuing
clusters are not self-similar fractals: the mean cluster width
perpendicular to the field direction grows faster 
than the height. The mean width exhibits power-law scaling with respect
to the cluster size, $\ave{r_x} \sim s^{l_x}$, $l_x = 0.61 \pm 0.01$,
but the mean height does not. The clusters tend 
to become elongated in the sedimentation direction and
the ratio of the single particle sedimentation velocity to single
particle diffusivity controls the degree of orientation. 
These results are obtained using a simulation method, which becomes
the more efficient the larger the moving clusters are. 

\end{abstract}


\maketitle

\section{Introduction} \label{intro_sec}

Aggregation of particles and particle clusters is still of great interest
not only due to the number of applications it has in chemical engineering, 
material sciences, and atmosphere research but also due to its
fundamental role as a simple model system for growth under
non-equilibrium
conditions~\cite{Meakin:PhysicaScripta46,Friedlander_book,Family_Landau}.
The effects of the interplay of diffusive
and ballistic motion in the case of fractal aggregates has been
considered only
recently~\cite{Reddy:JCIS82:1,Wang:JFluidMech295,Allain:PRL74,Senis:PRE55,Allain:JCIS178,Gonzalez:PRL86,Gonzalez:JPCM14,Leone:EPJE7,Odriozola:PRE67,Wu:L19}.
The relative strength of these two
mechanisms will vary with cluster size. 
This is the case, for example, in colloidal
suspensions, where both the diffusivity and the sedimentation velocity
are usually expected to depend algebraically on cluster size. In this
paper we study the effect of these two processes on the cluster
structure. 

The present understanding of colloidal aggregation under gravitation
is as follows. According to
experiments~\cite{Allain:JCIS178,Allain:PRL74} 
sedimenting clusters do not rotate, their anisotropy is
independent of their size, and they have no preferred orientation. 
Based on a scaling relationship between cluster size and sedimentation
velocity the sedimenting clusters are argued to be self-similar and
their fractal dimension to be
significantly larger than that in diffusion-limited
aggregation. The restructuring of particles inside a cluster caused by
hydrodynamic stresses was claimed to be the reason for sedimenting
clusters being more compact. Recent
simulations~\cite{Gonzalez:PRL86,Gonzalez:JPCM14} have 
shown that 
even without restructuring it is possible to produce clusters
with an apparent fractal dimension that is close to
the value observed in experiments,
if there is a velocity difference between clusters of different sizes.
The cross-over to the sedimentation-dominated regime can
be measured by directly observing the average size of the
aggregates; however in this regime the fractal dimension still
follows DLCA-like scaling \cite{Wu:L19}.

\begin{figure}
\includegraphics[width=\linewidth]{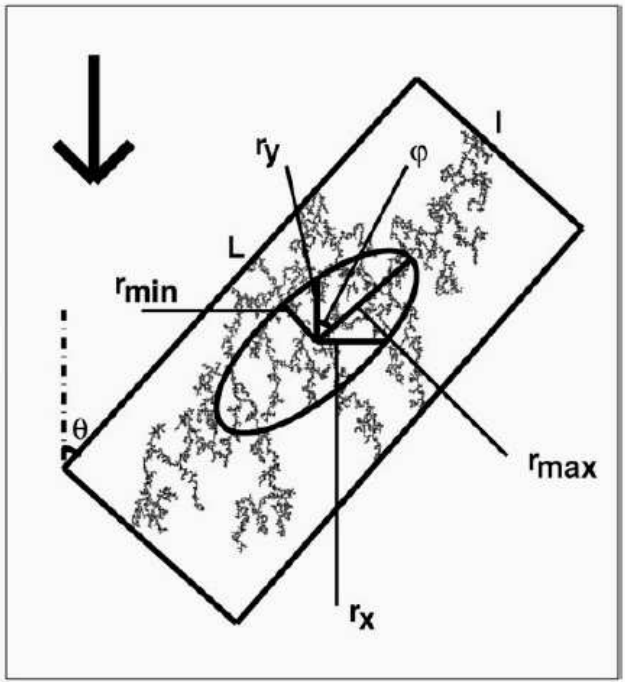}
\caption{
Illustration of the principal radii of gyration $r_{\rm min}$ and
$r_{\rm max}$ calculated from the radius of gyration matrix $T$
and the radii $r_x$ and $r_y$, which point perpendicular 
and parallel to the sedimentation velocity (shown by the thick arrow),
respectively. The longer 
side of the rectangle shown equals to the maximal distance between any
of the particles of the cluster and the angles $\varphi$ and $\theta$
are used to measure the orientation with respect to the direction of
the sedimentation velocity.}
\label{illustrationfig}
\end{figure}

In this article we study scaling properties of clusters, which are
formed in cluster-cluster aggregation when both the diffusivity and
the sedimentation velocity depend algebraically on cluster size. 
This induces a cross-over from diffusion-limited aggregation to 
a process where large clusters grow while settling by aggregating
smaller clusters from a time-dependent size distribution, $n_s(t)$.
We are mainly interested in generic features and hence concentrate
on two dimensions, 
in which the effects of sedimentation on cluster structure can be
studied with the most ease numerically. To compare our results to the
existing ones, we also consider other structural characteristics on a
qualitative level. For quantitative studies one should study more
elaborate models with hydrodynamics and other delicate issues
involved. The main focus here is on the
self-similarity of aggregates, which is the crucial assumption made in
the analysis of data in previous, similar studies. For this purpose,
we consider the scaling of four radii as a function of
cluster size. These are indicated in figure~\ref{illustrationfig}. 

Our main result is
that {\it sedimenting clusters are not self-similar
fractals}. 
Hence, an algebraic relationship - as assumed a priori here -
between cluster velocity and its mass
does not necessarily imply self-similarity of the aggregates.  
The mean
cluster width, considered in the direction perpendicular to the
sedimentation velocity, grows algebraically as a function of cluster
size but the mean cluster height does not follow a simple
power-law, at least not in the size range considered. 
It is an open question as to whether the asymptotic scaling of the
clusters transverse to the velocity
follows the mean cluster width, or, vice versa; much
larger simulations are needed to settle this issue.
Also the principal radii of gyration, given by the
eigenvalues of the radius of gyration matrix (see
Fig.~\ref{illustrationfig} and Eq.~\eqref{Tmatrixeq}),
grow with rates that are not equal. 
We present a scaling
argument that explains how the width grows algebraically
but simultaneously leaves the origin of the actual value of the exponent
open. The fact that this particular quantity exhibits
scaling brings up some interesting issues, like
where does the universal, Peclet-number independent value
originate from?

The difference in the growth rates 
indicates that clusters can not be described simply using the fractal
dimension and methods relying on isotropic scaling properties. For example,
studies using the radius of gyration give only an effective measure for
the cluster structure. Anisotropic growth also implies that the
cluster shape and its structure
change in time, and we consider the former in particular
in detail. We also analyze
the orientation of clusters and discuss the similarities and
differences between our results and previous experimental
and numerical studies of the same
phenomenon~\cite{Allain:JCIS178,Gonzalez:PRL86,Gonzalez:JPCM14}. 

This paper is organized as follows. 
The model is defined in section~\ref{model_sec}. A simulation method
becoming the efficient the larger clusters become is introduced in
section~\ref{compasp_sec}. The scaling properties using the radius of
gyration matrix are studied in
section~\ref{scalprop_sec}. Section~\ref{shape_sec} considers the
anisotropy and orientation of clusters. 
Section~\ref{conc_sec} concludes the paper. 

\section{Model} \label{model_sec}

Consider fractal clusters aggregating in a suspension. The sedimentation
caused by gravitation may be ignored for aggregates
with characteristic radii smaller than about 1~$\mu$m. In this region
the essential features of the 
growth are well described by the diffusion-limited cluster-cluster
aggregation~(DLCA)
model~\cite{Meakin:PhysicaScripta46,Meakin:PhaseTrans12}. The idea
is that clusters move diffusively,
and the  diffusion constant depends algebraically on
cluster size: $D(s) \sim s^\gamma$. For a
cluster diffusing in a quiescent fluid it can be argued that the 
diffusion exponent $\gamma = -1/\df$, where $\df$ is the
fractal dimension of the
cluster~\cite{Wiltzius:PRL58,Meakin:JCP82_3786_(1985),Wang:PRE60}.
Whenever clusters (particles are 
clusters of size one) collide, they irreversibly aggregate
together. No cluster restructuring is allowed. 

Both the dynamics and the structure of clusters formed in the DLCA are
well understood~\cite{Meakin:PhysicaScripta46,Meakin:PhaseTrans12}.  
Defining the number of clusters of size 
$s$ at time $t$ as $n_s(t)$, the cluster size distribution obeys dynamic
scaling $n_s(t) = S(t)^{-2} f(s/S(t))$, where the exponent
$-2$ follows from mass conservation and the average cluster
size $S(t) = \sum_s s^2n_s/\sum_s sn_s \sim t^z$ with $z$ being the
dynamic exponent. In two dimensions 
the fractal dimension of clusters is $\df =1.44 \pm
0.02$~\cite{Jullien:JPhysiqueLett45,Meakin:PL107A} 
and including rotational diffusion or varying the value of $\gamma$ has no
essential effect on it~\cite{Meakin:PhysicaScripta46,Meakin:PhaseTrans12}.
The anisotropy
measured using the ratio of the principal axis of gyration is about
$2.4$~\cite{Botet:JPA19}.

Here the DLCA model is considered in the presence of the
sedimentation of clusters. The force exerted on a cluster sedimenting
in a fluid consists of the buoyancy force $\vec{F}_b = V
\Delta \rho \vec{g}$ and 
the viscous drag force $\vec{F}_d = -C \vec{v}$, where $V$ is the
volume of the object (for an aggregate consisting of $s$ particles it
is $s$ times the volume of a single particle),
$\Delta \rho = \rho_p - \rho_f$ is the difference between particle and
fluid densities, $\vec{g}$ is the acceleration due to gravity, $\vec{v}$ is
the sedimentation velocity and $C$ is a positive constant depending on
the specific form of the cluster~\cite{Faber:book}. For a ball of
radius $r$ the constant 
$C=6\pi \eta r$, where $\eta$ is the kinematic viscosity. For complicated,
say fractal-like objects, $C$ is generally unknown and we make
the usual assumption that the aggregate will behave as a compact
object with an effective hydrodynamic radius $R_h$, i.e. $C \sim
R_h$~\cite{Meakin:JCP82_3786_(1985),Wiltzius:PRL58,VanSaarloos:Physica147A}.
If one further assumes a scaling relation between $R_h$ and $s$, the
sedimentation velocity depends 
algebraically on cluster size $v(s) \sim s^\delta$, where $\delta$ is
the sedimentation exponent.  
For a fractal one would have $s \sim R_h^{\df}$ resulting in  $\delta
= 1 - 1/\df$ but here we just take this as an assumption since it is
not a priori clear that the ensuing clusters are fractals with respect
to drag resistance. As $\gamma <0$ and
$\delta > 0$ aggregation is 
diffusion-limited for small clusters but becomes 
dominated by sedimentation for large ones. 

The model described above neglects many issues related to real
suspensions such as hydrodynamic interactions, restructuring and
rotation of clusters by diffusion or at aggregation to mention a
few. However, our purpose is not to try to simulate all aspects of
sedimentation driven aggregation but rather elucidate what kind of
universal behavior one could expect to have. Hence, the results
reported should be considered on a qualitative level when comparing to
experiments. We discuss further these issues and the assumptions
behind $\gamma$ and $\delta$ in the conclusions, in the light of the
simulation results obtained.

\begin{figure*}[ht]
\includegraphics[width=0.3\linewidth]{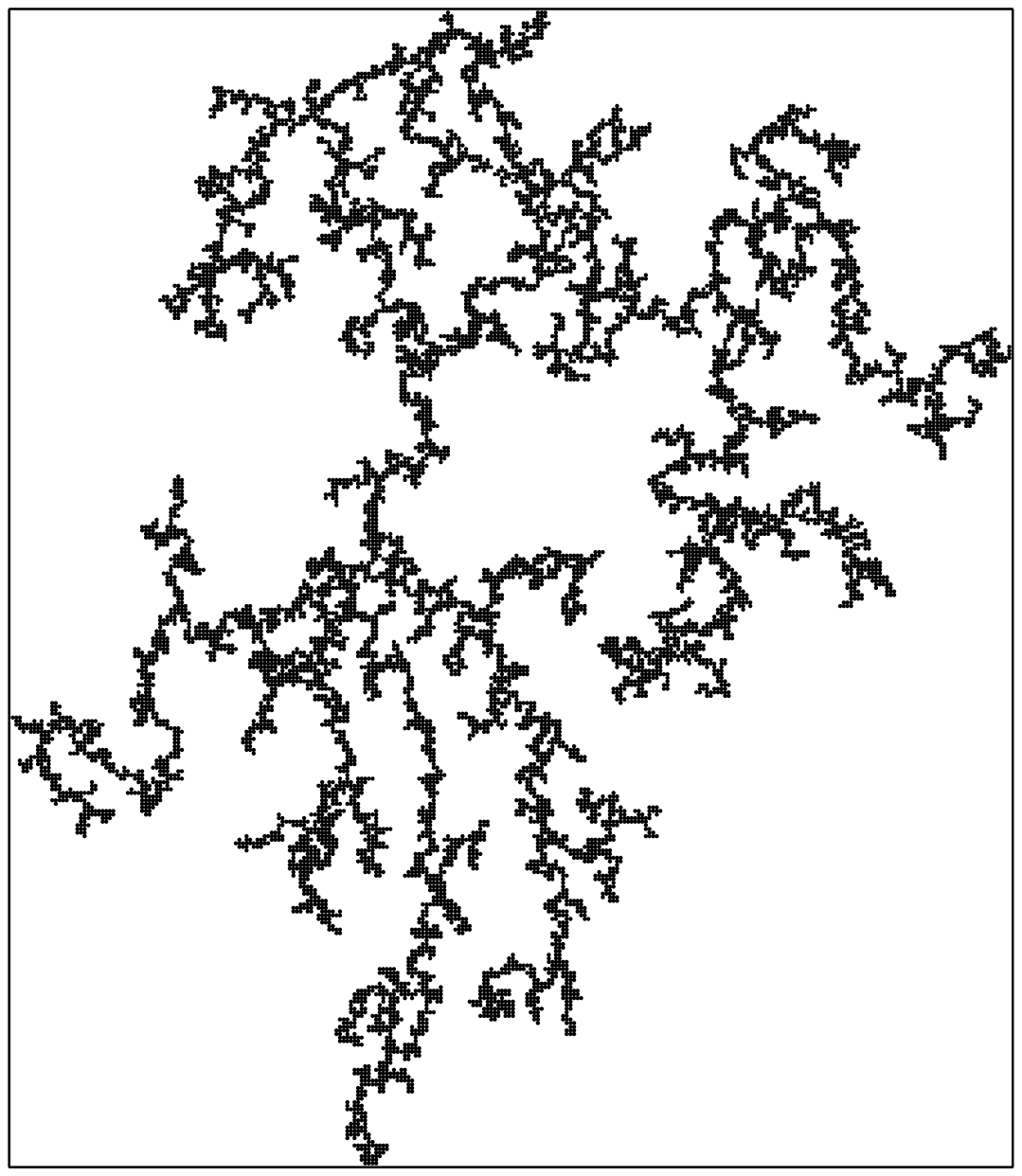}
\includegraphics[width=0.3\linewidth]{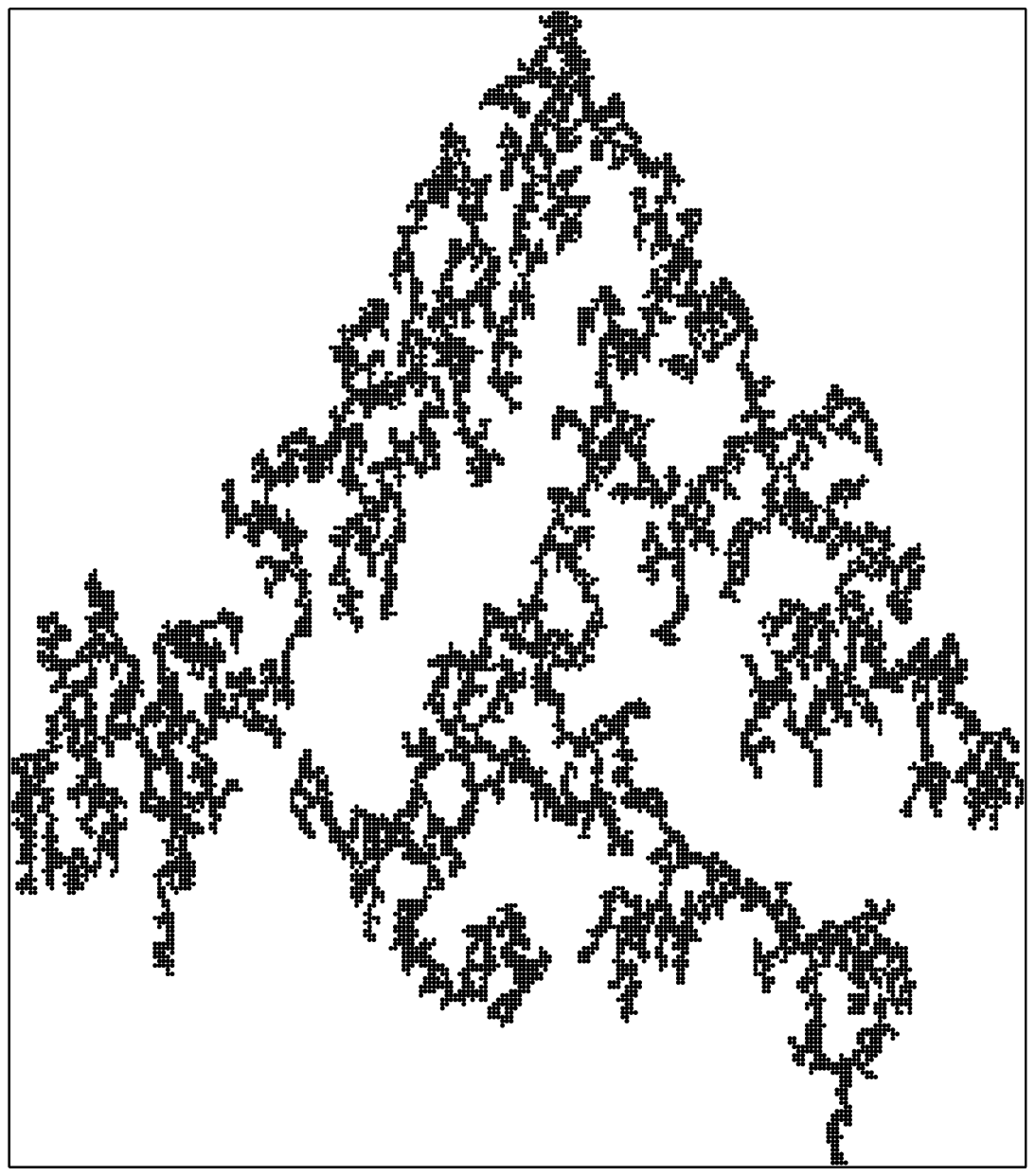}
\includegraphics[width=0.3\linewidth]{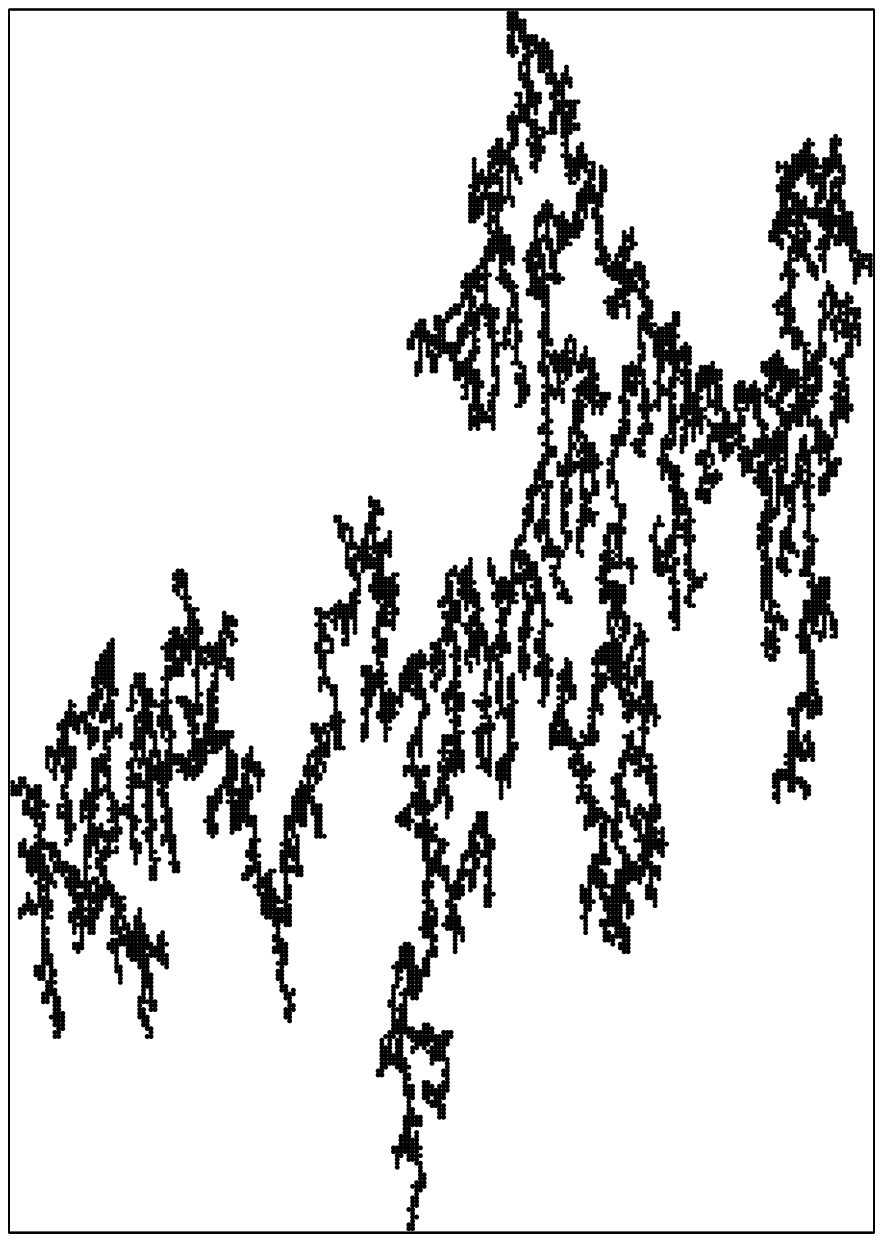}
\caption{Typical clusters formed in simulations for $\Pe = 0.01$,
1, and 10 (from left to right). They have 41167, 87585, and 80512
particles and are contained in rectangles of size 688 $\times$ 942,
1169 $\times$ 1307 and 1373 $\times$ 1625, respectively. The
sedimentation velocity points downwards.
}
\label{exampleclusters}
\end{figure*}

A useful measure for the relative strengths of diffusion and
sedimentation is the Peclet number~\cite{Ladd:PhysFluidsA5}
\begin{equation}
P_{\rm e}(s) = v(s) r / D(s),
\end{equation}
where $r$ is the radius of a single particle and $v = |\vec{v}|$. As
the Peclet number depends on cluster size, in the following we use the 
one particle Peclet number $\Pe = P_{\rm e}(1)$. 
Figure~\ref{exampleclusters} shows examples of clusters formed in
simulations for different values of $\Pe$.

\section{Computational Aspects} \label{compasp_sec}

The simulations are done on a two-dimensional lattice with periodic
boundary conditions. Initially a concentration $\phi$ of lattice sites
are filled randomly. One filled lattice site is considered as a single
particle. 
In the dynamics a particle or a cluster consisting of particles is selected
randomly and time is 
incremented by $N(t)^{-1} \Omega_{\rm max}^{-1}$, where $N(t)$ is the
number of clusters at time $t$ and $\Omega_{\rm max}$ is the maximum
mobility of any of the clusters at that time. The cluster mobility is
defined as $\Omega(s) = \pd(s) + \pu(s) + 2\pperp(s)$, where
$\pd(s)$, $\pu(s)$, and $\pperp(s)$ are proportional to the
probabilities for a cluster of size $s$ 
to move in the direction of the field, opposite to it and to the
directions perpendicular to it, respectively. 
The selected cluster (of size $s$)  
is moved if $x<\Omega(s)/\Omega_{\rm max}$, where $x \in (0,1)$ is a
random number selected from a uniform distribution. If the selected
cluster is not moved the simulation proceeds by selecting 
another cluster, otherwise another random number $y \in (0,1)$ is
selected from a uniform distribution. The cluster is
moved to the field direction if $y \le \pd(s)/\Omega(s)$, opposite to the
field direction if $\pd(s)/\Omega(s) < y \le
[\pd(s)+\pu(s)]/\Omega(s)$, and otherwise to one of the directions
perpendicular to the field with equal 
probabilities ($=\pperp(s)/\Omega(s)$). 
If after the move the cluster has no overlap with others the
move is accepted and another cluster is selected randomly. Otherwise
the move is taken back and the cluster is aggregated with all the
clusters it attempted to overlap before selecting a new cluster
for the next attempt.

The terms $\pd(s)$, $\pu(s)$, and $\pperp(s)$ are of the form
\begin{eqnarray}
\pd(s) & =  & \left(4 C_\parallel s^\gamma + C_\downarrow s^\delta +
C_\downarrow^2 s^{2\delta}\right)/2 \label{notsedincdiffeka} \nonumber
\\ 
\pu(s) & =  & \left(4 C_\parallel s^\gamma - C_\downarrow s^\delta + C_\downarrow
^2 s^{2\delta}\right)/2
\label{notsedincdiffvika} \\
\pperp(s) & =  & 2 C_\perp s^\gamma, \nonumber
\end{eqnarray}
where $C_\downarrow$, $C_\parallel$, and $C_\perp$ are non-negative
constants chosen such that $\pu(s) > 0$ for all cluster sizes. This
choice gives for the sedimentation velocity $v_s = 
C_\downarrow s^\delta$ and for the diffusivities $D_\parallel(s) =
C_\parallel s^\gamma$ and  $D_\perp(s) = C_\perp s^\gamma$, in the
directions parallel and 
perpendicular to the field, respectively. In the following we consider
only the case $C_\parallel=C_\perp$ and denote $D(s) = D_\parallel(s) =
D_\perp(s)$. Physically, only the ratio of the
sedimentation velocity to the diffusion constant is relevant and hence
we report the results as a function of the one particle Peclet number
$\Pe  = C_\downarrow/C_\perp$. 
The exponents are taken to be $\gamma = -0.70$ and $\delta = 0.30$,
which are obtained from relations $\delta = 1 + \gamma$ and $\gamma =
-1/\df$ using the fractal dimension of DLCA clusters $\df=1.44$. 
This is of course a valid choice initially, when DLCA is the
dominating process.
It should be underlined that the main point in these values
is that eventually sedimentation will prevail, and that the both
transfer mechanisms obey an algebraic dependence on the cluster size
$s$. 

It is known that in DLCA cluster structure, and especially the
effectuve fractal dimension, depends on
concentration~\cite{Lach-Hab:PRE54}.
In simulations reported 
here, the concentration is kept fixed to $\phi = 0.01$. The DLCA
calculations are done on a lattice of size $L_{\parallel} \times
L_{\perp} = 4000 \times 4000$ and when sedimentation
is included  $15000 \times 2000$ unless stated otherwise. 
To minimize finite size effects, all the simulations are run under the
conditions that none of the particles has traveled the size of
the lattice in the sedimentation direction and  the size of the
largest clusters is smaller than half of the lattice size in any
direction.  

The above choice for $\pd(s)$, $\pu(s)$, and $\pperp(s)$
assumes that sedimentation does not have any effect on the diffusion
properties of a cluster. This may not be the case as the velocity
fluctuations of sedimenting, compact, and non-aggregating particles
are larger in the sedimentation direction than perpendicular to 
it~\cite{Nicolai:PhysFluids7_3,Nicolai:PhysFluids7_12,Kalthoff:IntJofModPhysC7}.
The dependence of the strength of fluctuations on 
aggregate size is unknown, so we make the choice of isotropic
fluctuations. The precise form of fluctuations should be
unimportant as the upper critical dimension for sedimentation driven
aggregation is expected to be one~\cite{Hellen:PRE62}. 

Next we shortly discuss the implementation of the algorithm. It
is constructed to be efficient for low concentrations and for large
cluster sizes. Although the method is described using a
two-dimensional lattice, the generalization to higher dimensions and
the continuum case is straightforward. 

Each cluster is represented by a virtual rectangle, whose size is
the smallest possible one to enclose the cluster such that two sides
of the rectangle are parallel to the sedimentation velocity (see
Fig.~\ref{exampleclusters}). The sites belonging to a cluster are
stored with respect to the, say, lower left corner of the enclosing
rectangle, for two reasons. The first one is to minimize the time
needed to move a cluster. Regardless of the cluster size a move can be
performed by updating one single value i.e. either the $x$- or
$y$-coordinate of the lower left corner of the enclosing
rectangle. The second 
reason deals with the collisions of clusters. It is fast first to
check if the rectangles overlap each other and only if they do to
check for the overlap of the clusters. 

Obviously it would be very time-consuming to compare the rectangle
that is moved to all the other rectangles. Hence, the whole system is
divided into rectangular blocks and
clusters are identified to belong into blocks
according to the rectangles enclosing them. Then only 
clusters belonging to the same block(s) have to be checked.

\begin{figure}[h]
\includegraphics[width=\linewidth]{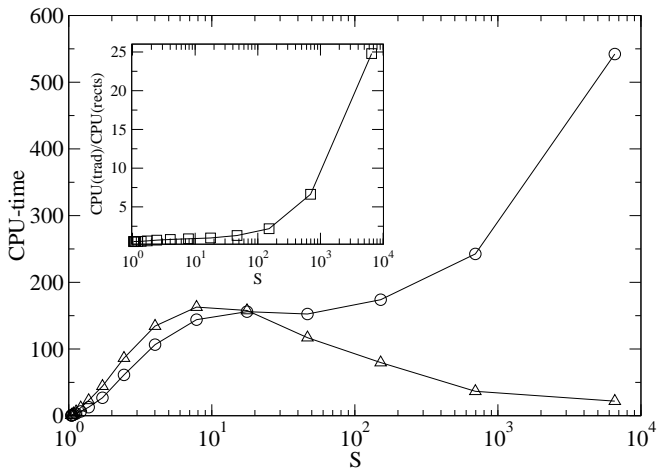}
\caption{CPU-time as a function of the average cluster size for
'traditional' ($\bigcirc$) and 'rectangle'-algorithms ($\triangle$). 
Each symbol represents the time used from the previous inspection time. 
The inset shows the ratio of 
the two CPU-times. 
}
\label{CPUtimecomparison}
\end{figure}

Figure~\ref{CPUtimecomparison} compares CPU-times using the
implementation described above to that with a ``traditional''
algorithm, where clusters are represented as occupied sites on a
lattice and moved site by site. For average cluster size smaller than
about ten the 
traditional algorithm is faster as one needs only to compare the occupancy
of a few sites. On the other hand, for large clusters the method
using enclosing rectangles becomes much faster, being independent
of cluster size, and usually only a couple of
comparisons are needed to check the overlap of a cluster
with others. In the ``traditional'' implementation one needs of
order $S(t)$ comparisons, which slows down the computation
considerably when $S(t)$ becomes large. 

In the
simulations reported here the ``traditional'' method is used for
$S(t)<10$ and the ``rectangle'' one for $S(t)>10$. 
Hence, most of the computation time
is used when the average cluster size is small. To obtain good
statistics from the interesting, large cluster size region, we further
apply a cloning method. It has proved to be efficient in many
applications including, for example, studies of percolation
clusters, native states of polymers, and
reaction-diffusion systems~\cite{Grassberger:COMPPHYSCOMM}. The
basic idea is simple: at  
a fixed average cluster size $S_{\rm copy}$ one makes $n_{\rm copy}$
copies of the cluster configuration and continues the Monte Carlo
simulation independently for each of these systems. We use $S_{\rm
copy} = n_{\rm copy}=10$. We tested that the results obtained
using copying were indistinguishable from the ones obtained without
it. The averages reported in this article are taken over $40$ runs without
copying and $10$ runs which were copied $10$ times. To increase the
scaling window further would necessitate enormous CPU resources;
even with the algorithm presented above we estimate that to augment
the maximum cluster size by one order of magnitude would require
of the order of one CPU-year on a fast computer.

Note that in the range of $s$-values shown, the typical scaling
of $S(t)$ is not algebraic manifesting
a cross-over from diffusive to sedimentation-dominated dynamics
\cite{Hellen:PRE62}. Since in the asymptotic regime the mean-field
dynamic exponent 
$z=1/(1-\delta-1/d_{\rm eff})$, for any reasonable value of $d_{\rm eff}$
(the effective fractal dimension) $z$ will have a value much
larger than unity, and the cross-over effects are noticeable
since the DLCA value is much smaller.

\section{Scaling properties of clusters} \label{scalprop_sec}

Complex aggregates are often characterized be their fractal dimension
$\df$, which can be found by the relation of the cluster size to its 
characteristic radius 
\begin{equation}
s \sim \langle R \rangle ^{\df}, 
\label{m_vs_R_relation}
\end{equation}
where the brackets denote averaging over clusters of size $s$. 
Usually one considers the radius of gyration 
\begin{equation}
R_g = \sqrt{\frac{1}{s}\sum_{i=1}^{s} \left( \vec{r}_i - \vec{r}_{\rm _{CM}} 
\right)^2}, \label{Rg_eq}
\end{equation}
where $\vec{r}_i$ denotes the position of the $i$th particle 
and $\vec{r}_{\rm _{CM}} = \frac{1}{s}\sum_{i=1}^s \vec{r}_i$ is the
center of mass of the cluster. 

As the field breaks the rotational symmetry it is not obvious that
clusters will scale isotropically. Hence, we
consider the radius of gyration matrix 
\begin{equation}
T=\left( 
\begin{array}{cc}
T_{xx} & T_{xy} \\
T_{yx} & T_{yy}
\end{array} \right) \, , \label{Tmatrixeq}
\end{equation}
where
\begin{equation}
T_{\alpha \beta} = \frac{1}{s}\sum_{i=1}^{s} 
(\vec{r}_{i,\alpha} - \vec{r}_{{\rm CM},\alpha})
(\vec{r}_{i,\beta} - \vec{r}_{{\rm CM},\beta})  \label{Rg_matrix_eq}
\end{equation}
and $x$ and $y$ refer to the $x$- and $y$-components of the position
vectors. The larger and smaller eigenvalues of $T$, denoted by
$\lambda_{\rm max}$ and $\lambda_{\rm min}$, respectively, are
related to the radius of gyration by $R_g^2 = \lambda_{\rm max} +
\lambda_{\rm min}$. The square roots of the
eigenvalues, called the principal radii 
of gyration, $r_{\rm max,min} = \lambda_{\rm max,min}^{1/2}$, may be
considered as lengths of an ellipsoid describing the cluster shape and
the corresponding 
normalized eigenvectors $\vec{e}_{\rm max}$ and $\vec{e}_{\rm min}$
define the directions of the principal axes. The 
preferred direction is given by the eigenvector corresponding to the
larger eigenvalue. We further consider the radii 
in the directions perpendicular and parallel to the field, 
which are related to the $T$
by the relations $r_x = T_{xx}^{1/2}$ and $r_y = T_{yy}^{1/2}$,
respectively, and to the radius of gyration by
$R_g^2=r_x^2+r_y^2$. 
Figure~\ref{illustrationfig} depicts the four radii for a cluster
formed with $\Pe=0.1$ and of size 90110.

\begin{figure}
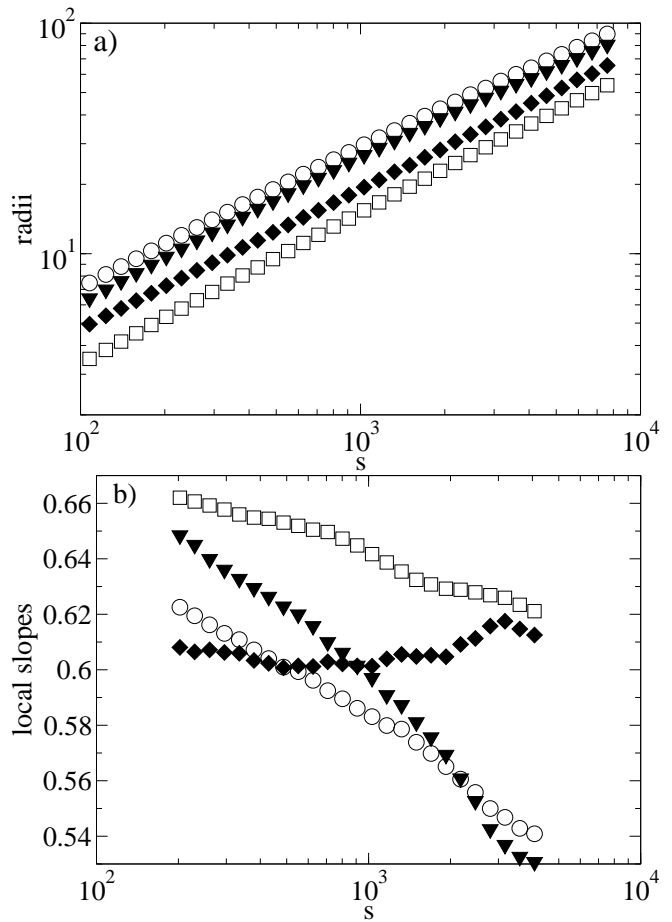

\includegraphics[width=\linewidth]{./radius1.eps} \\
\includegraphics[width=\linewidth]{./sliding1.eps}
\caption{(a) The mean radii $\ave{r_{\rm max}}$~($\bigcirc$), $\ave{r_{\rm
min}}$~($\square$), $\ave{r_x}$~($\blacklozenge$), and
$\ave{r_y}$~($\blacktriangledown$) as a function of the cluster size for $\Pe 
= 1$. (b) The local slopes. 
}
\label{radii_max_min_etc_pe1}
\end{figure}

\begin{figure}
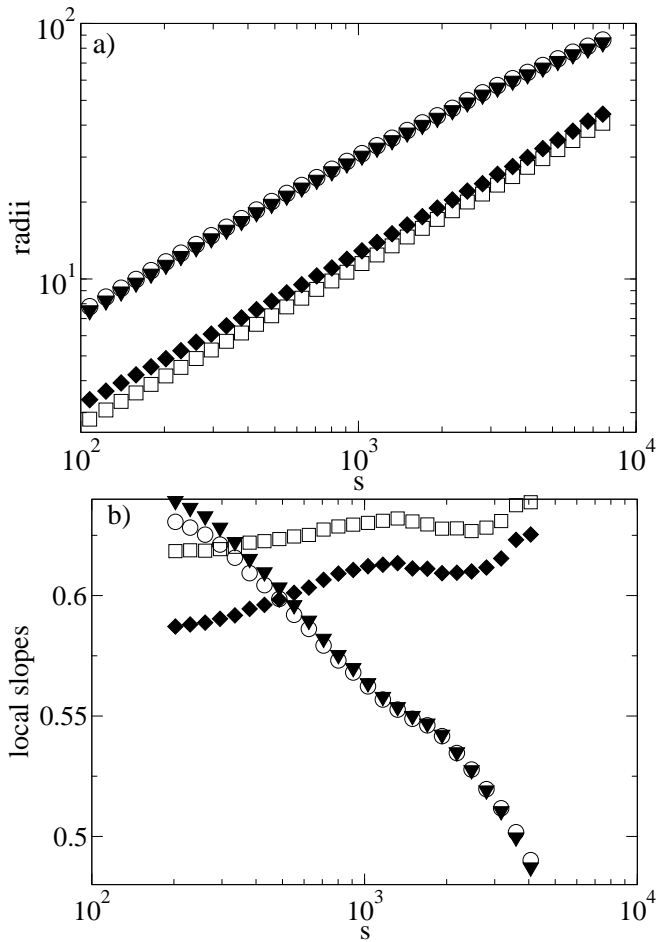

\includegraphics[width=\linewidth]{./radius10.eps} \\
\includegraphics[width=\linewidth]{./sliding10.eps}
\caption{(a) The mean radii $\ave{r_{\rm max}}$~($\bigcirc$), $\ave{r_{\rm
min}}$~($\square$), $\ave{r_x}$~($\blacklozenge$), and
$\ave{r_y}$~($\blacktriangledown$) as a function of the cluster size for $\Pe 
= 10$. (b) The local slopes. 
}
\label{radii_max_min_etc_pe10}
\end{figure}

Figures~\ref{radii_max_min_etc_pe1} (a) and 
\ref{radii_max_min_etc_pe10} (a) show the growth of the average radii
defined above as a function of the cluster size for $\Pe = 1$ and 
$\Pe = 10$, respectively. Although
the growth seems rather similar for different radii, 
the local slopes (shown in Figures 4 (b) and 5 (b)) change continuously with
the cluster size. 
Only the radius
perpendicular to the field obeys nice scaling of form 
$\ave{r_x} \sim s^{l_x}$ with $l_x = 0.61 \pm 0.01$.

\begin{figure}
\includegraphics[width=\linewidth]{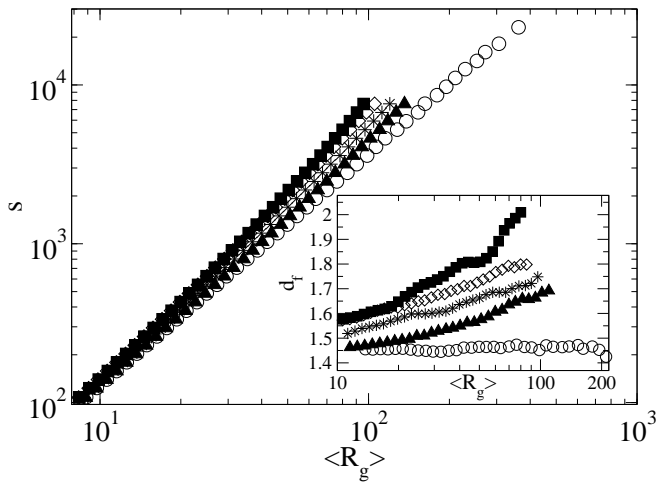}
\caption{Cluster size as a function of the radius of gyration for the
DLCA ($\bigcirc$), $\Pe=10$ ($\blacksquare$), $\Pe=1$ ($\diamondsuit$), 
$\Pe=0.1$ ($\star$), and $\Pe=0.01$ ($\blacktriangle$).
The inset shows the corresponding
local slopes indicating the change in the effective fractal dimension
of sedimenting clusters. }
\label{Rg_for_all}
\end{figure}

The overall behavior for other Peclet values ($\Pe$ = 0.01 and 0.1;
not shown) is rather similar 
and, especially, the value of $l_x$ is independent of the Peclet
number. For comparison, Figure~\ref{Rg_for_all} shows the scaling of the radius
of gyration [Eq.~$\eqref{Rg_eq}$] as function of cluster size for various 
Peclet numbers. The local slopes reveal that only for the DLCA there
is a scaling relation between these two. For a non-zero Peclet
number one has to consider the scaling using the radius of gyration
matrix [Eq.~\eqref{Tmatrixeq}] or some other observable not relaying
on isotropic scaling. One should emphasize that for all $\Pe > 0$
any asymptotic behavior is still far away in the range of accessible
radii of gyration; the actual values reflect more the cross-over
in the cluster shape and internal structure (density) than a tendency
to e.g. become compact in the limit $t\rightarrow \infty$. This can
be underlined by comparing with the scaling of the quantities,
in Figs.~\ref{radii_max_min_etc_pe1} and \ref{radii_max_min_etc_pe10}.

The reasons for the existence or lack of scaling
with respect to $r_x$ and $r_y$ can be considered
by a simple scaling argument. If $r_x$ scales with $s$, then also its
derivative with respect to $s$ 
does so. We have that
$\partial r_x/ \partial s \approx \Delta r_x / \Delta s$ and
\begin{equation}
\Delta r_x = P(r_x) \times \delta r_x
\end{equation}
where $P(r_x)$ denotes the probability that $r_x$ changes and $\delta
r_x$ measures the typical change in $r_x$ 
per such an aggregation event.

Consider a cluster of mass $s_1$, which collides
with another one, with $s_2 < s_1$. Then
$\delta r_x \sim r_x(s_2) \sim s_2^{l_x}$ and
$P(r_x) \sim r_x(s_2) / r_x (s_1) \sim (s_2/s_1)^{l_x}$.
Thus 
$\partial r_x / \partial s \sim s_2^{2 {l_x}}/(s_1^{l_x} s_2)$
and by making the ansatz $\langle s_2 \rangle \sim a s_1$ 
(since $s_2$ is cut-off by $s_1$, $a<1$),  we obtain that
\begin{equation}
\partial r_x / \partial  s \sim s_1^{{l_x}-1},
\end{equation}
exactly as one should if the growth is algebraic.

In the case of $r_y$, the probability $P(r_y)$
does not have to be algebraic. This is simply so because
(see Fig.~\ref{exampleclusters}) the roughness of the
lower part of a cluster growing via sedimentation will
also play a role: many of the smaller clusters will be
``swallowed'' inside the fjords opening downwards.
It seems feasible that this leads to a change in the
compactness of aggregates, which in turn results
in a non-algebraic relationship between $r_y$ and $s$.
This is in fact a ``restructuring''
process, related to the changing compactness of aggregates. 

Note the analogy with single
cluster growth in various ballistic growth models (with
shadowing, see Ch.~5.7.2 of \cite{Meakin:book}). There it is
known that the bulk of the cluster becomes compact ($\df = d$),
and that the boundary of the cluster undergoes interesting
roughening behavior. In our case, $l_x$ does not
show signs of such a cross-over (to compact geometry)
though $l_y$ might do so asymptotically.

\begin{figure}
\includegraphics[width=\linewidth]{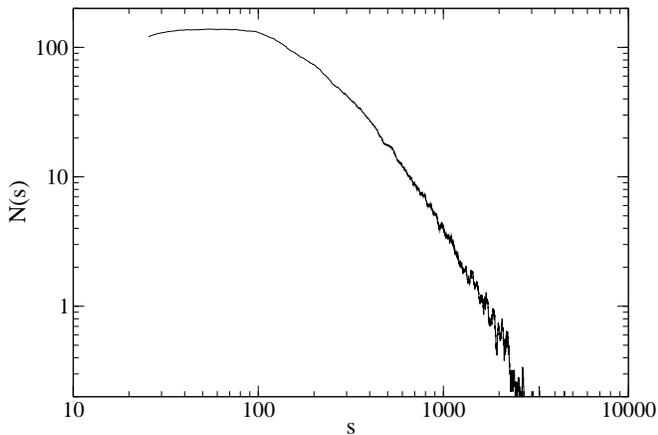}
\caption{The averaged (40 runs) cluster size distribution $n_s(t)$
for $\Pe= 1$, at $t=1024$.}
\label{nsigure}
\end{figure}

The equivalence to cluster growth is by no means clear in our case since 
the ``deposited'' clusters are sampled from the $n_s(t)$ at each $t$,
the distribution of which is broad (Fig.~\ref{nsigure} demonstrates an
example, see also ref.~\cite{odrio}). 
These clusters have a typical width-to-height ratio
depending on $s$, making the ``deposition problem'' much
more complicated than such ones usually are. This will 
remain true for any $t$, as the $n_s(t)$ remains non-trivial.
The implication is that for any finite Peclet number 
there should be a slow cross-over process that depends
self-consistently on the $n_s(t)$ and on the cluster structure
that has been formed through sampling smaller clusters from
the same. In any case, the
fact follows that the cluster structure is anisotropic:
the more recently aggregated parts of clusters are not 
scale-invariant with the parts that have been created earlier in the
aggregation process. 

Consider next the local exponents. For typical cluster sizes the local 
exponents of $\ave{r_{\rm min}(s)}$ and
$\ave{r_x(s)}$ seem to be close to the same value (see 
Figs.~\ref{radii_max_min_etc_pe1} (b) and
\ref{radii_max_min_etc_pe10} (b)). Also the slopes of
$\ave{r_{\rm max}(s)}$ and $\ave{r_y(s)}$ behave similarly for large
cluster sizes. This indicates that the cluster are elongated in the
field direction, i.e., $\ave{r_y} > \ave{r_x}$. The elongation is more
pronounced for high Peclet numbers. 
The same conclusion is obtained by directly considering the angular 
distribution using the eigenvector 
corresponding to the larger eigenvalue~(see Section~\ref{shape_sec}). 

Although in the size region studied $\ave{r_y} > \ave{r_x}$, the local
slope of $\ave{r_x}$ is larger than that of $\ave{r_y}$ indicating
that clusters grow faster in the direction 
perpendicular to the field. This implies that very large clusters
should become elongated perpendicular to the sedimentation
direction. Rough estimates for the crossover values are 
$s=8000$, 80000, and 300000 for $\Pe=0.1$, 1, and 10,
respectively, and the corresponding radii of gyration 80, 200 and 500.
These values are, unfortunately, larger than we are able 
to simulate but should be achievable in
experiments, see for example~\cite{Allain:JCIS178}. 
The most important conclusion, is however a choice between
two asymptotic scenarios. The data implies that asymptotically
$\ave{r_x(s)}$ will scale with the maximum radius. Hence, should
this be the case, there are two possiblities to begin with: either the 
maximum radius also starts to follow the $\ave{r_x(s)}$,
with its scaling exponent (0.61 above), or then the
asymptotic scaling of both could become volume-like ($\sim s^{0.5}$).

\section{Anisotropy and orientation} \label{shape_sec}

Next we consider how the cluster shape evolves with time. To have an
easily comparable measure for clusters of different 
sizes, we consider the scaled half-width
of a cluster as a function of its scaled height $Y = (y - y_{\rm
min})/(y_{\rm max}-y_{\rm min})$, where $y_{\rm max}$ and $y_{\rm min}$
are the maximum and minimum $y$-coordinates of any of the particles in
the cluster. The half-width at height $Y$ is defined as 
\begin{equation}
w(Y) = \frac{1}{2}\frac{x_{\rm max}(Y)-x_{\rm min}(Y)}{y_{\rm max}-y_{\rm min}},
\end{equation}
where $x_{\rm max}$ and $x_{\rm min}$
are the maximum and minimum $x$-coordinates of any of the particles at
that height. 

\begin{figure}
\includegraphics[width=0.6\linewidth]{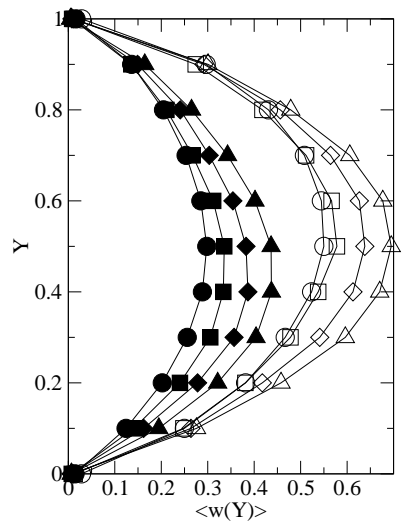}
\caption{Average cluster half-width as a function of scaled height for $\Pe
= 0.01$~(open symbols) and $10$~(filled symbols). The clusters are divided
into four size categories: $2500-5000$~($\bigcirc$),
$5000-7500$~($\square$), 7500-10000($\Diamond$),
$10000-\ldots$~($\triangle$).}
\label{widthfigure}
\end{figure}

The relative width of a cluster increases with size as can be seen
from Figure~\ref{widthfigure}, in which the half-width is shown
for two Peclet numbers. Note, that 
the relative width may well become larger than the height of a
cluster. 
For $\Pe = 0.01$ the clusters are symmetric with
respect to the point $Y=0.5$ but for $\Pe = 10$ they are wider at the bottom. 
This is due to the fact that as large
clusters sediment faster than small ones they gather mass at bottom
side. In other words, the upper part of a cluster is shielded by its
lower part. For large clusters this results in a 
triangular-like shape, a particularly nice example of which is shown
in Fig.~\ref{exampleclusters}. The sedimentation driven aggregates are
sparse and have  
complicated scaling properties (see Figs.~\ref{radii_max_min_etc_pe1}
and \ref{radii_max_min_etc_pe10}) as the 
clusters formed in the diffusion-dominated regime are themselves fractals. 
It is an open question as to what is the best characteristics
of this kind of anisotropic cluster shapes. One possibility would
be to look at the lower and upper parts separately (e.g. the widths
thereof, after splitting the cluster at the center of mass).

\begin{figure}
\includegraphics[width=\linewidth]{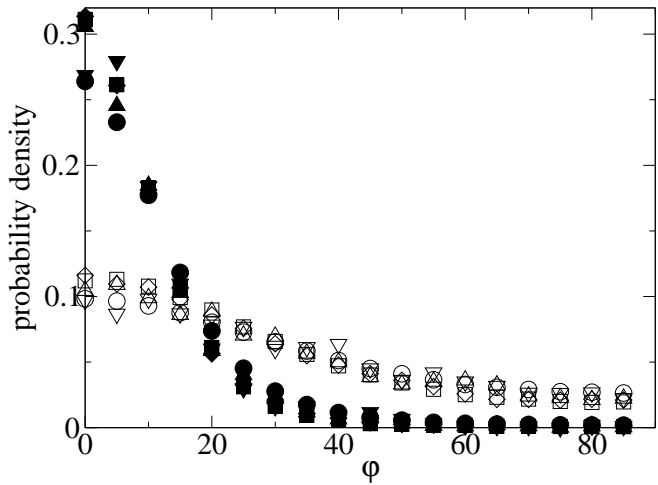}
\caption{Angular distribution with respect to the sedimentation
direction  for
$\Pe=1$~(open symbols) and $\Pe=10$~(filled symbols)
and for several size classes: $1000-2500$ ($\bigcirc$),
2500-5000 ($\square$), 5000-7500
($\diamondsuit$), 7500 - 10000 ($\triangle$) and 10000 $\ldots$
($\triangledown$).
}
\label{angular_dist}
\end{figure}

The orientation of sedimenting clusters is considered using the
angle $\varphi = {\rm arccos}(\vec{e}_y \cdot \vec{e}_{\rm max})$ between the
eigenvector corresponding to the larger eigenvalue and the unit vector
in the direction of the external field. Figure~\ref{angular_dist}
shows the angular distribution for two values of the Peclet
number. 

To study the effect of cluster size on orientation the
clusters are divided in five different size classes. There is no
difference in orientation with respect to cluster size. The main
point, however, is that the clusters are orientated such that
they prefer to have the longer principal axis aligned with the
field and that the orientation distribution depends on the Peclet
number. For example, for $\Pe = 10$, there are practically no clusters 
with $\varphi > 45^o$. This is not in contrast to the
conclusion made using the half-width as there is an important
difference between these two approaches, when one averages them over
cluster sizes. The averaging of the principal radii gives
information about the typical cluster shape whereas averaging cluster
width takes into account the selection of a preferred direction. We
believe, that if one could simulate larger cluster sizes, 
another peak would start to grow at $\varphi = 90^o$. 

To consider our simulations in the light of the three-dimensional experiments 
 we also calculate cluster anisotropy and orientation as in
Ref.~\cite{Allain:JCIS178}. First, a cluster is enclosed in 
a rectangle, whose longer sides give the maximal
distance between any of the points in the cluster and the shorter ones
just touch the cluster (see Fig.~\ref{illustrationfig}).
The cluster ``radius'' is defined as
$R=(L+l)/4$, where $L$ ($l$) denotes the length of the longer
(shorter) edge. The orientation is measured using the angle
$\theta$ between the sedimentation velocity and the longer edge.

\begin{figure}
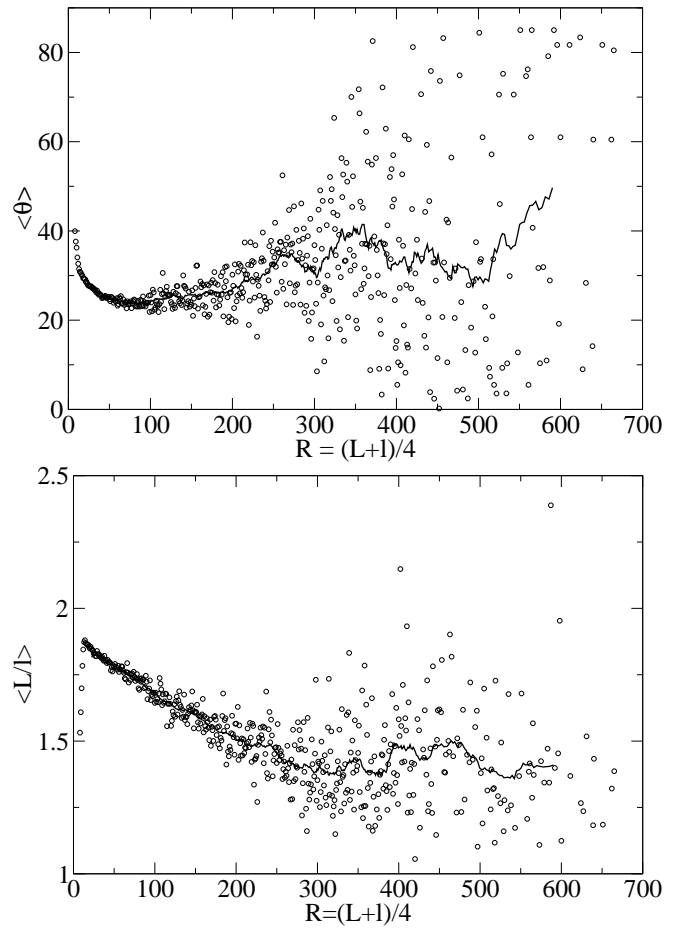

\includegraphics[width=\linewidth]{./maxbox1_angle.eps}
 \includegraphics[width=\linewidth]{./maxbox1_ratio.eps}
\caption{a) Average orientation angle $\ave{\theta}$ and b) aspect
ratio $\ave{L/l}$ as a function of radius $R=(L+l)/4$ for $\Pe =
1$. Solid lines represent sliding averages using 30 data points. 
}
\label{Allain_kuvat}
\end{figure}

Figure~\ref{Allain_kuvat} shows the average anisotropy and the average
orientation angle as a function of $R$. The averages are taken over
intervals of length one, which explains the noisiness of the data for
large cluster sizes. There is a
clear transient region, where both quantities 
vary with cluster size but for large clusters they become constant within
statistical errors and the value of the anisotropy ratio $L/l \approx
1.4$. This is the same as the value of three-dimensional
clusters~\cite{Allain:JCIS178}. However, since $r_x$ and
$r_y$ have different scaling properties this would be a bit surprising,
if it were the asymptotic value. It seems most likely that
the constant value reflects more the fact that the statistics
becomes sparse, with only a few clusters in this 
range of $R$.  One can compare the anisotropy to the ratio of $r_{max}$
and $r_{min}$; this starts for $s$ small from the DCLA value and is thus
different.
Another explanation might be the crossover from field-elongated
aggregates to ones that are wider in the direction perpendicular
to the sedimentation velocity: 
At $R=300$ the cluster size $s$ is approximately 45000, 
which is of the same order
of magnitude than the value estimated for the crossover 
$\langle r_x \rangle \approx \langle r_y \rangle$ in Sec.~\ref{scalprop_sec} 
($s$=80000).

\section{Conclusions} \label{conc_sec}

We have studied the structure of two-dimensional
clusters, which aggregate at contact and whose motion is dominated by
diffusion for small cluster sizes and by sedimentation for large
ones. Considering the eigenvalues of the radius of gyration matrix, we
show that due to the sedimentation the aggregates grow slower
in the direction of the sedimentation velocity than perpendicular to
it. Moreover, only the cluster width scales with cluster size and
hence the clusters are not self-affine either, at the very least
within the size range currently accessible in simulations. 
We underline that the data implies a cross-over scale, at
which the maximal cluster radius will begin to scale with
the average cluster width, but no further conclusions can
be drawn about this regime.

Interestingly, the scaling exponent
characterizing the growth of the width is independent of the Peclet
number, which characterizes the strength of the influence of
sedimentation on aggregation. It is the one that defines
``self-similarity'' in the growing aggregates, but only in a limited
sense since it concerns only a one-dimensional characteristics of 
the clusters (in contrast to e.g.\ a fractal dimension).
The width scaling argument presented leaves open {\em what} the value of the
width exponent would be but hints about the reasons as to why scaling
is obtained in this respect, only. Here the clear conclusion is that
since the Peclet number has no influence on the value ($\approx 0.61$)
it appears to be a universal one, and as such should be measured
experimentally.  The value to be obtained from three-dimensional
experiments would hence be independent of the viscosity of the
suspending liquid but not necessarily take the same value as obtained
here. Recent experiments have shown that the radius of gyration
can be used to distinguish between the DLCA and sedimentation
regimes, with a clear change in the rate of increase with time \cite{Wu:L19}.
However, the fractal dimension was shown to initially remain
at the corresponding DLCA value. We believe this underlines
the fact that in such simulations as ours it is easier to concentrate
on measures that characterize the behavior of the large aggregates,
as for instance $R_g(s)$.

Let us next discuss the justification of the simulation rules with the
asymmetric growth in mind. As far as the growth is dominated by
diffusion, the aggregates will be self-similar and diffusion
characterized through $D(s) \sim s^{\gamma}$ is reasonable. When  
sedimentation starts to dominate this no longer holds but then the
precise characterization of diffusion is unimportant anyway. In
retrospect the algebraic dependence of the 
sedimentation velocity on the cluster size is just a lucky choice,
a posteriori justified by the fact that $r_x$ scales algebraically. On
the other hand, the relation $\delta = 1-1/d_f$ is invalid. This
raises the academic question that how does the value of the universal
exponent $l_x$ depend on the values of $\gamma$ and $\delta$? One may
argue as follows. As the fractal dimension of DLCA clusters is
independent of $\gamma$, it probably does not change the scaling
properties of sedimenting clusters either. However, $\delta$ may and
probably will have an effect; at least on the universality grounds
there is no reason why the value of $l_x$ would not change with
$\delta$.  It would of course be possible to attempt much more
elaborate and time-consuming  simulations, in which the hydrodynamic radius 
of each aggregate is computed self-consistently.

The anisotropic growth is also seen when studying the
average form of aggregates. Larger clusters are always relatively
wider than their smaller counterparts. For small Peclet numbers the
clusters are rather elliptic but they became wider at the ``bottom part''
due to a shielding effect for large Peclet numbers. 
We also studied the orientation of clusters with respect to the
direction of the sedimentation velocity. Clusters are
oriented such that their height is larger than the width,
but eventually the
situation should be the opposite as the cluster width grows faster than
the height. The larger the Peclet number the stronger the cluster
orientation anisotropy.

The inequality of the growth rates in different direction implies that
the scaling properties 
of sedimenting clusters can not be studied using the radius of
gyration, which implicitly assumes the clusters to scale similarly
in different directions. As there exist data from 
experiments~\cite{Allain:JCIS178} and 
simulations~\cite{Gonzalez:PRL86,Gonzalez:JPCM14} 
considering aggregation of sedimenting clusters, it would be
worthwhile to check for the possibility of anisotropic growth in these
cases, too. Experiments or simulations with larger aggregates could also
reveal if the height of sedimenting aggregates will eventually scale
with their size. It would definitely be also
of interest to make simulations in the three dimensional case, and if
cluster restructuring or rotational diffusion is allowed.

{\bf Acknowledgments} - The authors  
thank the Academy of Finland, Center of Excellence program,
for financial support. 
E.~K.~O.~H. further thanks Jenny and Antti Wihuri Foundation for 
financial support, while M.~J.~A would like to acknowledge
the SMC Center, Universit\'a
``La Sapienza'', Rome, for hospitality and prof. Joachim Krug
for a reminder.

\bibliographystyle{prsty}

\end{document}